\documentclass{article}
\usepackage{spconf}
\usepackage[dvips]{graphicx}
\usepackage{amsmath}
\usepackage{amssymb}
\usepackage{citesort}
\usepackage{times}
\usepackage{amsfonts}
\usepackage{float}
\usepackage{amsthm}
\usepackage{color}
\usepackage{psfrag}
\usepackage{setspace}

\newtheorem{proposition}{Proposition}

\ninept

\begin{document}

\title{Jamming resistant receivers for Massive MIMO}

\name{Tan Tai Do, Emil Bj\"ornson, and Erik G. Larsson \thanks{The authors would like to acknowledge useful discussions on this topic with Mohammad Razavizadeh.}}
\address{Department of Electrical Engineering (ISY), Link\"oping University (LiU), Sweden}

\maketitle
\vspace{-0.2cm}
\begin{abstract}
We design jamming resistant receivers to enhance the robustness of a massive MIMO uplink channel against jamming. In the pilot phase, we estimate not only the desired channel, but also the jamming channel by exploiting purposely unused pilot sequences. The jamming channel estimate is used to construct the linear receive filter to reduce impact that jamming has on the achievable rates. The performance of the proposed scheme is analytically and numerically evaluated. These results show that the proposed scheme greatly improves the rates, as compared to conventional receivers. Moreover, the proposed schemes still work well with stronger jamming power.
\end{abstract}


\vspace{-0.2cm}
\section{Introduction}
As a promising candidate for the emerging 5G wireless communication networks \cite{Mar10TWC,NLM13TCOM},
massive multiple-input multiple-output (MIMO) has recently received a lot of research attention. This technology has demonstrated unprecedented spectral efficiencies by serving many tens of users on the same time-frequency resource.

The physical layer security aspects of massive MIMO are relatively unexplored. While massive MIMO is robust against passive eavesdropping \cite{RZR15CM}, active jamming attacks is an issue. The extra pilot contamination caused by jamming leads to a significant performance loss \cite{BKA15CNS}.
Although jamming exists and has been identified as a critical problem for reliable communications, there are only a few works focusing on the jamming aspects in massive MIMO \cite{RZR15CM,BKA15CNS,PRB16WCL,WLW15TWC}.
For instance, the authors of \cite{RZR15CM,BKA15CNS} consider secure transmission in a downlink massive MIMO, in the presence of attackers capable of jamming and eavesdropping. Optimized jamming is considered for uplink massive MIMO in \cite{PRB16WCL}, which shows that a smart jammer can cause substantial jamming pilot contamination that degrades the sum rate. The paper \cite{WLW15TWC} investigates the artificial noise-aided jamming design for a massive MIMO transmitter in Rician fading channels.

Pilot contamination appears when the pilot signal, transmitted for estimation of a user channel, is interfered by another signal \cite{Mar10TWC}. The typical effect is that the base station (BS) cannot use the estimated channel to coherently combine the desired signal, without also coherently combining the interference. Pilot contamination between legitimate users of the system is a challenge in massive MIMO, but can be handled by pilot coordination across cells \cite{BLD16TWC} or by exploiting spatial correlation \cite{Yin2013a,Bjornson2017a}. The problem with jamming pilot contamination is more difficult to deal with, because the jammer attempts to create maximum pilot contamination rather than minimum. Since the structure and properties of the jamming attack is limited, a typical approach to deal with jamming signals is to treat them as additive noise and design the transceivers as if there was no jamming \cite{BKA15CNS,PRB16WCL}. However, jamming is not noise-like since the desired channel estimate is correlated with the jamming channel.

In this work, we propose jamming resistant receivers to achieve robustness of the massive MIMO uplink against jamming attacks. In particular, we construct the receive filters using not only the desired channel estimate but also an estimate of the jamming channel. To this end, we exploit purposely unused pilot sequences, which are orthogonal to the pilot sequences assigned to the users, to estimate the jamming channel. The estimate of the jamming channel is used to design receive filters that reject the jamming signal. We consider two different receive filters, which are motivated by the conventional linear minimum mean square error (MMSE) and zero-forcing (ZF) filters. In order to evaluate the performance of the proposed schemes, the achievable rates are analyzed and closed-form large-scale approximations are obtained. Simulation results are also provided to verify our analysis.

\vspace{-4mm}

\section{Problem Setup}
\begin{figure}[t]
\centering
\psfrag{BS}[][][0.9]{$\mathrm{Base~station}$}
\psfrag{US}[][][0.9]{$\mathrm{User}$}
\psfrag{JM}[][][0.9]{$\mathrm{Jammer}$}
\psfrag{gu}[][][0.9]{$\mathbf{h}$}
\psfrag{gj}[][][0.9]{$\mathbf{g}$}
\includegraphics[width=4.5cm]{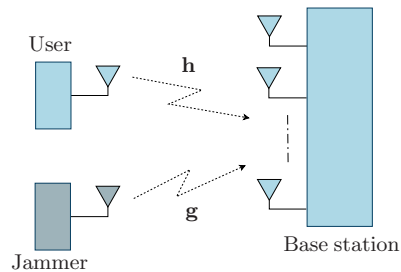}
\caption{Massive MIMO uplink under a jamming
attack.}\label{fig:system_model} \vspace{-4mm}
\end{figure}

We consider a single-user massive MIMO uplink consisting of a BS, a legitimate user and a jammer as depicted in Fig.~\ref{fig:system_model}. We assume that the BS is equipped with $M$ antennas, while the legitimate user and the jammer have a single antenna each. This basic model captures the main principle of jamming, and can be easily generalized to having multiple legitimate users.

Let us denote $\mathbf{h}\in \mathbb{C}^{M\times 1}$ and $\mathbf{g}\in \mathbb{C}^{M\times 1}$ as the channel vectors from the legitimate user and the jammer to the BS, respectively. We assume that the elements of $\mathbf{h}$ are independent and identically distributed (i.i.d.) zero-mean circularly-symmetric complex Gaussian (ZMCSCG) random variables, i.e., $\mathbf{h}\sim \mathcal{CN}(0,\beta_\mathrm{u}\mathbf{I}_M)$, where $\beta_\mathrm{u}$ represents the large-scale fading. Similarly, we assume $\mathbf{g}\sim \mathcal{CN}(0,\beta_\mathrm{j}\mathbf{I}_M)$, where $\beta_\mathrm{j}$ is the large-scale fading. The channels $\mathbf{h}$ and $\mathbf{g}$ are independent.

We consider a block-fading model, in which the channel remains constant during a coherence block of $T$ symbols, and varies independently from one coherence block to the next. The communication between the legitimate user and the BS follows a two-phase transmission protocol. In the first phase (pilot phase), the legitimate user sends pilot sequences to the BS for channel estimation. In the second phase (data transmission phase), the legitimate user transmits the payload data to the BS. We assume that the jammer attacks the system both in the pilot and data transmission phases.
\vspace{-0.2cm}
\subsection{Pilot Phase}
During the first $\tau$ symbols of a coherence block ($\tau<T$), the user transmits a pilot sequence $\mathbf{s}_\mathrm{u}$ of length $\tau$ symbols. This pilot originates from a pilot codebook  $\mathcal{S}$ containing $\tau$ orthogonal unit-norm vectors. We assume that $\tau$ is larger than the number of active users, which in this cases means that $\tau\geq 2$. We further assume that the legitimate system uses a pilot hopping scheme such that the jammer cannot know the user's current pilot sequence. Therefore, the jammer randomly chooses a jamming sequence $\mathbf{s}_{\mathrm{j}}$ uniformly distributed over the unit sphere. By sending the jamming sequence $\mathbf{s}_{\mathrm{j}}\in \mathbb{C}^{\tau \times 1}$, which satisfies $\|\mathbf{s}_{\mathrm{j}}\|^2=1$, the jammer hopes to interfere with the channel estimation.

Accordingly, the received signal at $M$ antennas of the BS in $\tau$ symbol times of the pilot phase can be stacked and given by
\begin{align}
\label{ytma}
\mathbf{Y}_\mathrm{t}=\sqrt{\tau p_\mathrm{t}}\mathbf{h}\mathbf{s}_\mathrm{u}^T+\sqrt{\tau q_\mathrm{t}}\mathbf{g}\mathbf{s}_\mathrm{j}^T+\mathbf{N}_\mathrm{t},
\end{align}
where $\mathbf{Y}_\mathrm{t}\in \mathbb{C}^{M \times \tau}$, $p_\mathrm{t}$ and $q_\mathrm{t}$ are the transmit powers of the user and jammer during the pilot phase, respectively. The additive noise matrix $\mathbf{N}_\mathrm{t} \in \mathbb{C}^{M \times \tau}$ is assumed to have i.i.d.~$\mathcal{CN}(0,1)$ elements.

\vspace{-0.2cm}
\subsection{Data Transmission Phase}
During the last $(T-\tau)$ symbols of a coherence block, the user transmits its
payload data to the BS and the jammer continues to interfere by sending a jamming signal. Let us denote by $x_\mathrm{u}$
($\mathbb{E}\{|x_\mathrm{u}|^2\}=1$) and $x_\mathrm{j}$
($\mathbb{E}\{|x_\mathrm{j}|^2\}=1$) the transmitted signals
from the user and the jammer, respectively. The received $M \times 1$ signal at the BS is
\begin{align}
\label{yd}
\mathbf{y}_\mathrm{d}=\sqrt{p_\mathrm{d}}\mathbf{h}x_\mathrm{u}+\sqrt{q_\mathrm{d}}\mathbf{g}x_\mathrm{j}+\mathbf{n}_\mathrm{d},
\end{align}
where $p_\mathrm{d}$ and $q_\mathrm{d}$ are the transmit powers of the user and the jammer in the data transmission phase, respectively. The noise vector  $\mathbf{n}_\mathrm{d}$ is assumed to have i.i.d.~$\mathcal{CN}(0,1)$ elements.

To detect $x_\mathrm{u}$, the BS uses a linear receive filter as follows:
\begin{align}
\label{y}
y=\mathbf{a}^H\mathbf{y}_\mathrm{d}=\sqrt{p_\mathrm{d}}\mathbf{a}^H\mathbf{h}x_\mathrm{u}+
\sqrt{q_\mathrm{d}}\mathbf{a}^H\mathbf{g}x_\mathrm{j}+\mathbf{a}^H\mathbf{n}_\mathrm{d},
\end{align}
where $\mathbf{a}\in \mathbb{C}^{M \times 1}$ is the receive filter, which will be carefully selected in the next section to reject the jamming.

The received signal in (\ref{y}) can be rewritten as
\begin{align}
\nonumber
y=\sqrt{p_\mathrm{d}}\mathbb{E}\{\mathbf{a}^H\mathbf{h}|\mathbf{s}_\mathrm{j}\}x_\mathrm{u}+\sqrt{p_\mathrm{d}}(\mathbf{a}^H\mathbf{h}-
\mathbb{E}\{\mathbf{a}^H\mathbf{h}|\mathbf{s}_\mathrm{j}\})x_\mathrm{u}+&\\
\label{y2}
\sqrt{q_\mathrm{d}}\mathbf{a}^H\mathbf{g}x_\mathrm{j}+\mathbf{a}^H\mathbf{n}_\mathrm{d}.&
\end{align}
By treating $\sqrt{p_\mathrm{d}}\mathbb{E}\{\mathbf{a}^H\mathbf{h}|\mathbf{s}_\mathrm{j}\}$ as the deterministic channel that the desired signal is received over and treating the last three terms (which are uncorrelated with $x_\mathrm{u}$) as worst-case independent Gaussian noise, an achievable rate for the massive MIMO uplink is
\begin{align}
\label{R}
R= \left(1-\frac{\tau}{T}\right) \mathbb{E}_{\mathbf{s}_\mathrm{j}}\left\{\log_2\left(1+\rho\right)\right\},
\end{align}
\vspace{-0.2cm}
where
\vspace{-0.1cm}
\begin{align}
\label{rho}
\rho=\frac{p_\mathrm{d}|\mathbb{E}\{\mathbf{a}^H\mathbf{h}|\mathbf{s}_\mathrm{j}\}|^2}
{p_\mathrm{d}\texttt{var}\{\mathbf{a}^H\mathbf{h}|\mathbf{s}_\mathrm{j}\}
+q_\mathrm{d}\mathbb{E}\{|\mathbf{a}^H\mathbf{g}|^2|\mathbf{s}_\mathrm{j}\}+\mathbb{E}\{\|\mathbf{a}\|^2|\mathbf{s}_\mathrm{j}\}}
\end{align}
is the effective signal-to-interference-and-noise ratio (SINR).

We note that the effective SINR $\rho$ in (\ref{rho}) is conditioned on $\mathbf{s}_\mathrm{j}$. In order to realize the achievable rate in (\ref{R}), the BS needs to know the numerator and the denominator of $\rho$. Although $\mathbf{s}_\mathrm{j}$ is assumed to be unknown, we will later show that $\rho$ only depends on the correlations of $\mathbf{s}_\mathrm{j}$ and the legitimate pilot sequences, which can be estimated with high accuracy based on the received pilot signal's power when $M$ is large. We also stress that the receiver processing proposed in the next section will not exploit any instantaneous knowledge of $\mathbf{s}_\mathrm{j}$.

\vspace{-0.3cm}
\section{Channel estimation and jamming resistant filter design}
The achievable rate in (\ref{R}) highly depends on choice of receive filter $\mathbf{a}$. In this section, we consider two different receive filters, which are constructed based on not only the estimate of the desired channel but also on an estimate of the jamming channel.

\vspace{-0.4cm}
\subsection{Channel Estimation}
\vspace{-0.1cm}
In order to estimate the desired channel $\mathbf{h}$, the received pilot signal $\mathbf{Y}_\mathrm{t}$ is first correlated with the user's pilot sequence $\mathbf{s}_\mathrm{u}$ as
\begin{align}
\label{yth}
\mathbf{y}_\mathrm{t}=\mathbf{Y}_\mathrm{t}\mathbf{s}_\mathrm{u}^*=\sqrt{\tau p_\mathrm{t}}\mathbf{h}+\sqrt{\tau q_\mathrm{t}}\mathbf{s}_\mathrm{j}^T\mathbf{s}_\mathrm{u}^*\mathbf{g}+ \mathbf{N}_\mathrm{t}\mathbf{s}_\mathrm{u}^*.
\end{align}
The linear MMSE estimate of $\mathbf{h}$ given $\mathbf{y}_\mathrm{t}$ is given by \cite{Kay93}
\begin{align}
\label{he}
\mathbf{\widehat{h}}=c_\mathrm{u}\mathbf{y}_\mathrm{t}\triangleq\alpha_1 \mathbf{h}+\alpha_2 \mathbf{g}+ \mathbf{n}_1,
 \end{align}
where $c_\mathrm{u}=\frac{\sqrt{\tau p_\mathrm{t}}\beta_\mathrm{u}}{\tau p_\mathrm{t} \beta_\mathrm{u}+q_\mathrm{t}\beta_\mathrm{j}+1}$, $\alpha_1=c_\mathrm{u}\sqrt{\tau p_\mathrm{t}}$, $\alpha_2=c_\mathrm{u}\sqrt{\tau q_\mathrm{t}}\mathbf{s}_\mathrm{j}^T\mathbf{s}_\mathrm{u}^*$, and $\mathbf{n}_1 \sim \mathcal{CN}(0,c_\mathrm{u}^2\mathbf{I}_M)$.

As we can see from (\ref{he}), the desired channel estimate $\mathbf{\widehat{h}}$ is correlated with the jamming channel $\mathbf{g}$, i.e., $\mathbb{E}\{ \mathbf{g}^H \mathbf{\widehat{h}}|\mathbf{s}_\mathrm{j} \} = M\alpha_2\beta_j$.
Without the knowledge of the jamming channel $\mathbf{g}$, the receive filter $\mathbf{a}$ is generally chosen as a linear function of $\mathbf{\widehat{h}}$. One example is maximal ratio combining (MRC) with  $\mathbf{a} = \mathbf{\widehat{h}}$. Any such receive filter, which is correlated with the jamming channel, also amplifies the jamming signal and thus degrades the rate \cite{BKA15CNS}.

In order to mitigate the effect of the jamming, we propose to design the receive filter based on both $\mathbf{h}$ and $\mathbf{g}$. However, since $\mathbf{h}$ and $\mathbf{g}$ are not available at the BS, we can construct receive filters using their estimates instead. Recall that there is at least one unused pilot sequence in the system, which is orthogonal to the user's pilot $\mathbf{s}_\mathrm{u}$. By projecting the received pilot signal $\mathbf{Y}_\mathrm{t}$ onto this unused pilot sequence, the user's pilot signal is eliminated, leaving only the jamming signal (and noise). The resulting signal is
\begin{align}
\label{ge}
\mathbf{\widehat{g}}=\mathbf{Y}_\mathrm{t}\mathbf{s}_\mathrm{\overline{u}}^*\triangleq b \mathbf{g}+ \mathbf{n}_2,
\end{align}
where $\mathbf{s}_\mathrm{\overline{u}}$ is the unused pilot sequence, $\mathbf{s}_\mathrm{u}^T\mathbf{s}_\mathrm{\overline{u}}^*=0$, $b=\sqrt{\tau q_\mathrm{t}}\mathbf{s}_\mathrm{j}^T\mathbf{s}_\mathrm{\overline{u}}^*$, and $\mathbf{n}_2 \sim \mathcal{CN}(0,\mathbf{I}_M)$. This is an estimate of the jamming channel.

\begin{figure*}[t]
\begin{align} \tag{11}
\label{rho_mmse}
\rho_{\mathrm{mmse}} \asymp \frac{Mp_\mathrm{d}\alpha_1^2\beta_\mathrm{u}^2}{M\left(\frac{\sigma/(q_\mathrm{d}M)+1}{\sigma/(q_\mathrm{d}M)+
\gamma_\mathrm{j}}\right)^2q_\mathrm{d}|\alpha_2|^2\beta_\mathrm{j}^2+\left(\alpha_1^2\beta_\mathrm{u}(p_\mathrm{d}\beta_\mathrm{u}+1)
+c_\mathrm{u}^2+|\alpha_2|^2\beta_\mathrm{j}\frac{|b|^2\beta_\mathrm{j}+(\sigma/(q_\mathrm{d}M)+1)^2 }{(\sigma/(q_\mathrm{d}M)+\gamma_\mathrm{j})^2}\right)}
\end{align} \vspace{-7mm}
\end{figure*}

\begin{figure*}[t]
\begin{align} \tag{13}
\label{rho_zf}
\rho_\mathrm{zf} \asymp \frac{Mp_\mathrm{d}\alpha_1^2\beta_\mathrm{u}^2}{Mq_\mathrm{d}|\alpha_2|^2\frac{\beta_\mathrm{j}^2}{\gamma_\mathrm{j}^2}+
\left(\alpha_1^2\beta_\mathrm{u}(p_\mathrm{d}\beta_\mathrm{u}+1)+c_\mathrm{u}^2+
|\alpha_2|^2\frac{\beta_\mathrm{j}}{\gamma_\mathrm{j}}\right)}
\end{align}
\hrule
\end{figure*}

Based on the estimates $\mathbf{\widehat{h}}$ and $\mathbf{\widehat{g}}$, we will construct two receive filters, which are inspired by the conventional MMSE and ZF filters.

\vspace{-0.35cm}
\subsection{MMSE-type Receive Filter}
First, we consider the MMSE receive filter, which is optimal when the receiver has perfect CSI. We rewrite the received signal in (\ref{yd}) as
\begin{align}
\nonumber
\mathbf{y}_\mathrm{d}=\sqrt{p_\mathrm{d}}\mathbf{\widehat{h}}x_\mathrm{u}+\sqrt{q_\mathrm{d}}\mathbf{\widehat{g}}x_\mathrm{j}+
\sqrt{p_\mathrm{d}}\mathbf{e}_\mathrm{u}x_\mathrm{u}+\sqrt{q_\mathrm{d}}\mathbf{e}_\mathrm{j}x_\mathrm{j}+ \mathbf{n}_\mathrm{d},
\end{align}
where $\mathbf{e}_\mathrm{u}\triangleq \mathbf{h}-\mathbf{\widehat{h}}$ and $\mathbf{e}_\mathrm{j}\triangleq \mathbf{g}-\mathbf{\widehat{g}}$ are the desired and jamming channel estimation errors, respectively. By treating $\mathbf{w}\triangleq\sqrt{p_\mathrm{d}}\mathbf{e}_\mathrm{u}x_\mathrm{u}+\sqrt{q_\mathrm{d}}\mathbf{e}_\mathrm{j}x_\mathrm{j}+ \mathbf{n}_\mathrm{d}$ as equivalent uncorrelated additive Gaussian noise, an MMSE-type of receive filter can be obtained as
\begin{align}
\label{ammse}
\mathbf{a}_{\mathrm{mmse}}=\left(\mathbf{\widehat{g}}\mathbf{\widehat{g}}^H+\frac{\Psi}{q_\mathrm{d}}\right)^{-1}\mathbf{\widehat{h}},
\end{align}
where $\Psi$ is the covariance matrix of the signal associated with estimation errors plus noise, i.e.,
\begin{align}
\nonumber
\Psi=\mathbb{E}\{\mathbf{w}\mathbf{w}^H\}=\sigma\mathbf{I}_M,
\end{align}
where $\sigma=p_\mathrm{d}\beta_\mathrm{u}(1-c_\mathrm{u}\sqrt{\tau p_\mathrm{t}})+q_\mathrm{d}(\beta_\mathrm{j}(1+q_\mathrm{t})+1)+1$.

Note that our setup includes jamming pilot contamination, which makes the equivalent noise $\mathbf{w}$ correlated with the estimated channels $\mathbf{\widehat{h}}$ and $\mathbf{\widehat{g}}$. The receive filter in (\ref{ammse}) is thus no longer a true MMSE filter, i.e., $\mathbf{a}_{\mathrm{mmse}}$ may not be optimal. Thus, we call $\mathbf{a}_{\mathrm{mmse}}$  an ``MMSE-type'' receive filter.

Moreover, when $\mathbf{a}=\mathbf{a}_{\mathrm{mmse}}$, the effective received SINR in (\ref{rho}) can be evaluated as in the following proposition.

\begin{proposition}\label{Prop:MMSE-type}
Assume that the MMSE-type receive filter is used, then a large-scale approximation of the effective SINR is given in (\ref{rho_mmse}), where $\gamma_\mathrm{j}=|b|^2\beta_\mathrm{j}+1$ and $\asymp$ denotes asymptotic equivalent relation, i.e., $f_1[M]\asymp f_2[M]$ is equivalent to $f_1[M]-f_2[M]\mathop \to \limits^{M\to\infty}0$.
\end{proposition}

\setcounter{equation}{11}

The proof of Proposition \ref{Prop:MMSE-type} is given in the appendix. From \eqref{rho_mmse}, we can see that when $M\to\infty$, the effective received SINR converges to a finite limit. However, as we will show in the numerical results, the proposed receive filters can remarkably reduce the effect of the jamming attack.
\vspace{-0.3cm}
\subsection{ZF-type Receive Filter}
Although the MMSE-type receive filter in (\ref{ammse}) is a linear filter, it still has high complexity, especially in large systems since it involves a matrix inversion operation. We next consider a simpler receive filter, which does not require the matrix inversion operation. Motivated from the fact that the jamming signal is the main source of interference, we consider a receive filter that focuses on nulling the jamming signal, i.e., a ZF-type of receive filter.

Based on the estimated channels $\mathbf{\widehat{h}}$ and $\mathbf{\widehat{g}}$, a ZF-type receive filter can be obtained as
\vspace{-0.2cm}
\begin{align}
\label{azf}
\mathbf{a}_{\mathrm{zf}}=\left(\mathbf{I}_M-\frac{\mathbf{\widehat{g}}\mathbf{\widehat{g}}^H}{\|\mathbf{\widehat{g}}\|^2}\right)
\mathbf{\widehat{h}},
\end{align}
which projects $\mathbf{\widehat{h}}$ orthogonally to $\mathbf{\widehat{g}}$.
When $\mathbf{a}=\mathbf{a}_{\mathrm{zf}}$, the effective received SINR in (\ref{rho}) can be evaluated as in the following proposition.

\begin{proposition}\label{Prop:ZF-type}
Assume that the ZF-type receive filter is used, then a large-scale approximation of the effective SINR is given in (\ref{rho_zf}).
\end{proposition}

\setcounter{equation}{13}

The proof of Proposition \ref{Prop:ZF-type} is similar to the proof for Proposition~\ref{Prop:MMSE-type}, which is omitted for brevity.
We note that when the number of antennas $M\to\infty$, both the received SINRs with MMSE-type and ZF-type receive filters converge to the same finite value, i.e.,
\begin{align}
\lim_{M\to\infty}\rho_{\mathrm{mmse}}=\lim_{M\to\infty}\rho_{\mathrm{zf}}\triangleq \rho_{\mathrm{asy}}=
\frac{p_\mathrm{d}\alpha_1^2\beta_\mathrm{u}^2\gamma_\mathrm{j}^2}{q_\mathrm{d}|\alpha_2|^2\beta_\mathrm{j}^2}.
\end{align}
\vspace{-0.4cm}
\section{Numerical Results}
\vspace{-0.2cm}
In this section, we numerically evaluate the achievable rates of the massive MIMO uplink with different receive filters, including the proposed jamming resistant receivers. We consider a coherence block of $T=200$ symbols, $\tau=3$ and $\beta_\mathrm{u}=\beta_\mathrm{j}=1$. For comparison, we also include the rate achieved by the traditional MRC receiver, that does not use the estimate $\mathbf{\widehat{g}}$, i.e., $\mathbf{a}=\mathbf{a}_{\mathrm{mrc}}=\mathbf{\widehat{h}}$ \cite{PRB16WCL}.

\begin{figure}[t]
\centering
\includegraphics[width=8cm]{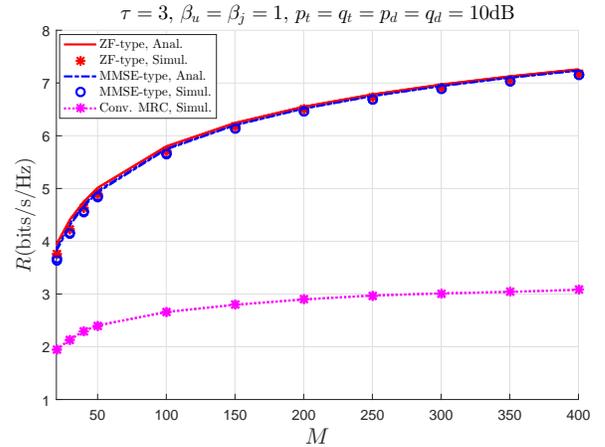}
\caption{Achievable rates for varying number of antennas.}
\label{vsM}
\end{figure}

Fig.~\ref{vsM} shows the achievable rates versus the number of antennas at the BS. As expected, the proposed receive filters based on the jamming channel estimate can remarkably improve the performance of the system, as compared to MRC. The achievable rates from our analysis (curves with ``Anal.'') are close to the Monte-Carlo simulations (curves with ``Simul.''), and will be asymptotically equal. Moreover, we can see that the ZF-type receive filter works particularly well and outperforms the MMSE-type receive filter. This shows that in massive MIMO, where the jamming effect is critical, a favorable receiver solution is to focus on nulling the jamming signal.

\begin{figure}[t]
\centering
\includegraphics[width=8cm]{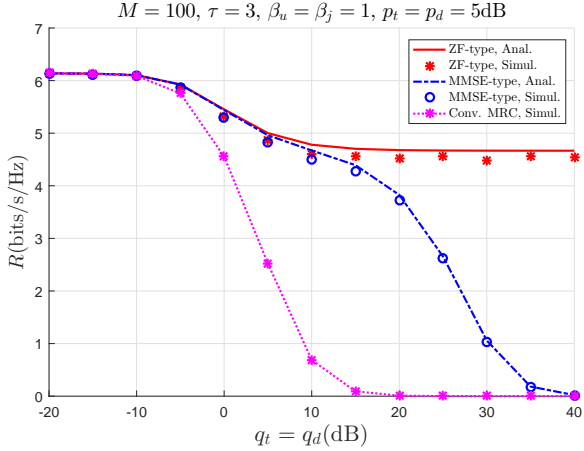}
\caption{Achievable rates for varying jamming powers.}
\label{vsqt} \vspace{-4mm}
\end{figure}

Next, we exemplify the effect of the jamming pilot power on the system performance. Fig.~\ref{vsqt} illustrates the achievable rates according to the values of jamming attack power $q_\mathrm{t}$. We fix the user's transmit powers as $p_\mathrm{t}=p_\mathrm{d}=5~\mathrm{ dB}$, and vary the jammer's transmit powers $q_\mathrm{t}=q_\mathrm{d}$ from $-20\,\mathrm{dB}$ to $40\, \mathrm{dB}$. As expected, the achievable rate with the MRC decreases with the increase of the jamming powers. However, it is interesting to see that the proposed schemes, especially with ZF-type receive filter, still work well with stronger jamming attacks. This can be explained by the fact that the proposed receive filters are constructed using the estimates of both the desired channel $\mathbf{h}$ and the jamming channel $\mathbf{g}$. When the pilot jamming power $q_\mathrm{t}$ increases, it does not only degrade the quality of the desired channel estimation but also improves the quality of the jamming channel estimation. Thus, the proposed receive filters can still work well if the improvement of the quality of $\mathbf{\widehat{g}}$ overcomes the degradation on the quality of $\mathbf{\widehat{h}}$. Moreover, we can analytically observe that the effective SINR $\rho_\mathrm{zf}$ in (\ref{rho_zf}) converges to a non-zero value when the data jamming power $q_\mathrm{d}$ tends to infinity as long as the pilot jamming power $q_\mathrm{t}$ grows with the same order. The detailed analysis is omitted here for the brevity.

\vspace{-0.2cm}
\section{Conclusion}
\vspace{-0.2cm}

A new jamming resistant receiver approach has been proposed to enhance the robustness of the massive MIMO uplink against jamming attacks. By exploiting purposely unused pilot sequences, the jamming channel can be estimated using the received pilot signal. The results show that the proposed receive filters, which were constructed using the jamming channel estimate, can greatly reduce the effect of jamming attack and improve the system performance. Moreover, the proposed schemes still work well when the jamming powers are large. Due to the critical effect of the jamming signal, a ZF-type receive filter, which focuses on nulling the jamming signal is a favorable solution. However, due to the presence of the jamming pilot contamination, the optimal receive filter is an open problem that is left for future research.
\vspace{-0.2cm}
\section{Appendix -- Proof of Proposition 1}\label{Pro1Proof}
\vspace{-0.2cm}
When the MMSE-type receive filter is used, i.e., $\mathbf{a}=\mathbf{a}_{\mathrm{mmse}}=\left(\mathbf{\widehat{g}}\mathbf{\widehat{g}}^H+
\frac{\sigma}{q_\mathrm{d}}\mathbf{I}_M\right)^{-1}\mathbf{\widehat{h}}$, the terms in (\ref{rho}) can be calculated as follows.\\
\vspace{0.9cm}
{\Large$\cdot$} \emph{The desired signal term $p_\mathrm{d}|\mathbb{E}\{\mathbf{a}^H\mathbf{h}|\mathbf{s}_\mathrm{j}\}|^2$}
\vspace{-0.7cm}

Let us consider
\begin{align}
\nonumber
\frac{\mathbf{a}^H\mathbf{h}}{M}&=\frac{\mathbf{\widehat{h}}^H}{M}\left(\mathbf{\widehat{g}}\mathbf{\widehat{g}}^H+
\frac{\sigma}{q_\mathrm{d}}\mathbf{I}_M\right)^{-1}\mathbf{h}\\
\nonumber
&=\frac{\mathbf{\widehat{h}}^H}{M}\frac{q_\mathrm{d}}{\sigma}\left(\mathbf{I}_M-
\frac{\mathbf{\widehat{g}}\mathbf{\widehat{g}}^H}{\sigma/q_\mathrm{d}+\|\mathbf{\widehat{g}}\|^2}\right)\mathbf{h}\\
\label{ah}
&=\frac{q_\mathrm{d}}{\sigma}\frac{\mathbf{\widehat{h}}^H\mathbf{h}}{M}-\frac{q_\mathrm{d}}{\sigma}
\frac{\frac{\mathbf{\widehat{h}}^H\mathbf{\widehat{g}}}{M}\frac{\mathbf{\widehat{g}}^H\mathbf{h}}{M}}
{\frac{\sigma}{Mq_\mathrm{d}}+\frac{\|\mathbf{\widehat{g}}\|^2}{M}},
\end{align}
where the second equality follows from the matrix inversion lemma. Since $\mathbf{h}$, $\mathbf{g}$, $\mathbf{n}_1$ and $\mathbf{n}_2$ are all independent, we have the following large-scale approximations that are tight when $M\to\infty$:
\begin{align}
\label{whh}
\frac{\mathbf{\widehat{h}^H}\mathbf{h}}{M}&=\frac{(\alpha_1 \mathbf{h}+\alpha_2 \mathbf{g}+\mathbf{n}_1)^H\mathbf{h}}{M}\asymp\alpha_1\beta_{\mathrm{u}},\\
\label{whwg}
\frac{\mathbf{\widehat{h}}^H\mathbf{\widehat{g}}}{M}&=\frac{(\alpha_1 \mathbf{h}+\alpha_2 \mathbf{g}+\mathbf{n}_1)^H(b\mathbf{g}+\mathbf{n}_2)}{M}\asymp\alpha_2^*b\beta_{\mathrm{j}},\\
\label{wgh}
\frac{\mathbf{\widehat{g}}^H\mathbf{h}}{M}&=\frac{(b\mathbf{g}+\mathbf{n}_2)^H\mathbf{h}}{M}\asymp0,\\
\label{wgwg}
\frac{\|\mathbf{\widehat{g}}\|^2}{M}&\asymp \gamma_{\mathrm{j}}.
\end{align}
Thus, it follows that $
\frac{\mathbf{a}^H\mathbf{h}}{M}\asymp\frac{q_\mathrm{d}}{\sigma}\alpha_1\beta_{\mathrm{u}}
$ and
\begin{align}
\label{t1}
\frac{p_\mathrm{d}|\mathbb{E}\{\mathbf{a}^H\mathbf{h}|\mathbf{s}_\mathrm{j}\}|^2}{M^2}\asymp p_\mathrm{d}\frac{q_\mathrm{d}^2}{\sigma^2}\alpha_1^2\beta_{\mathrm{u}}^2.
\end{align}
\vspace{0.7cm}
{\Large$\cdot$} \emph{The signal gain uncertainty term} $p_\mathrm{d}\texttt{var}\{\mathbf{a}^H\mathbf{h}|\mathbf{s}_\mathrm{j}\}$
\vspace{-0.4cm}

By using the large-scale approximations in (\ref{ah}), (\ref{whwg}), (\ref{wgh}), and (\ref{wgwg}) it follows that
\begin{align}
\nonumber
\frac{\mathbf{a}^H\mathbf{h}}{M}\asymp\frac{q_\mathrm{d}\alpha_1}{\sigma}\frac{\mathbf{h}^H\mathbf{h}}{M}.
\end{align}
Thus,
\begin{align}
\nonumber
\texttt{var}\left\{\frac{\mathbf{a}^H\mathbf{h}}{M}\bigg|\mathbf{s}_\mathrm{j}\right\}&\asymp\frac{q_\mathrm{d}^2\alpha_1^2}{\sigma^2M^2}
\texttt{var}\{\mathbf{h}^H\mathbf{h}\}\\
\nonumber
&=\frac{q_\mathrm{d}^2\alpha_1^2}{\sigma^2M^2}M\beta_\mathrm{u}^2=\frac{q_\mathrm{d}^2\alpha_1^2}{\sigma^2M}\beta_\mathrm{u}^2.
\end{align}
Therefore,
\begin{align}
\label{t2}
\frac{p_\mathrm{d}\texttt{var}\{\mathbf{a}^H\mathbf{h}|\mathbf{s}_\mathrm{j}\}}{M^2}\asymp p_\mathrm{d}\frac{q_\mathrm{d}^2\alpha_1^2}{\sigma^2M}\beta_\mathrm{u}^2.
\end{align}
\vspace{0.7cm}
{\Large$\cdot$} \emph{The jamming term $q_\mathrm{d}\mathbb{E}\{|\mathbf{a}^H\mathbf{g}|^2|\mathbf{s}_\mathrm{j}\}$}
\vspace{-0.4cm}

By following similar steps as for the desired signal term, when $M\to\infty$ we have
\begin{align}
\nonumber
\frac{\mathbf{a}^H\mathbf{g}}{M}\asymp\frac{q_\mathrm{d}}{\sigma}\alpha_2^*\beta_{\mathrm{j}}\frac{\sigma/(Mq_\mathrm{d})+1}
{\sigma/(Mq_\mathrm{d})+\gamma_j}
\end{align}
and
\begin{align}
\label{t3}
\frac{q_\mathrm{d}\mathbb{E}\{|\mathbf{a}^H\mathbf{g}|^2|\mathbf{s}_\mathrm{j}\}}{M^2}\asymp \frac{q_\mathrm{d}^3}{\sigma^2}\left(\frac{\sigma/(q_\mathrm{d}M)+1}{\sigma/(q_\mathrm{d}M)+
\gamma_\mathrm{j}}\right)^2|\alpha_2|^2\beta_\mathrm{j}^2.
\end{align}
\vspace{0.7cm}
{\Large$\cdot$} \emph{The noise term $\mathbb{E}\{\|\mathbf{a}\|^2|\mathbf{s}_\mathrm{j}\}$}
\vspace{-0.4cm}

Once again, by following similar steps as for the desired signal term, when $M\to\infty$ we have
\begin{align}
\label{t4}
\frac{\mathbb{E}\{\|\mathbf{a}\|^2|\mathbf{s}_\mathrm{j}\}}{M}\!\asymp\!
\frac{q_\mathrm{d}^2}{\sigma^2}\!\left(\!\alpha_1^2\beta_\mathrm{u}\!+\!c_\mathrm{u}^2\!+\!|\alpha_2|^2\beta_\mathrm{j}\frac{|b|^2\!\beta_\mathrm{j}
\!+\!(\sigma/(q_\mathrm{d}M)\!+\!1)^2 }{(\sigma/(q_\mathrm{d}M)\!+\!\gamma_\mathrm{j})^2}\!\right).
\end{align}
Substituting \eqref{t1}, \eqref{t2}, \eqref{t3}, and \eqref{t4} into
\eqref{rho} we obtain \eqref{rho_mmse}.
\bibliographystyle{IEEEtran}
\bibliography{IEEEabrv,Treferences}

\end{document}